\documentclass[12pt]{iopart}

\expandafter\let\csname equation*\endcsname\relax
\expandafter\let\csname endequation*\endcsname\relax
\usepackage{amsmath}
\usepackage{hyperref}
\usepackage{cleveref}
\crefname{equation}{equation}{equations}
\usepackage{xcolor}
\usepackage{dsfont}
\usepackage{empheq}
\usepackage{url}

\begin{document}

\title[]{Preserving the Hermiticity of the One-Body Density Matrix for a Non-Interacting Fermi Gas}%Input short title in [] space if needed

\author{L.M. Farrell$^{1,2}$\footnote[7]{Author to whom any correspondence should be addressed.}, D. Eaton$^{2,3}$, P. Chitnelawong$^{2,4}$, K. Bencheikh$^{5}$, and B.P. van Zyl$^{2}$\footnote[1]{The author has since passed.}
}

\address{$^1$ Department of Physics and Astronomy, McMaster University, Hamilton, Ontario, L8S 4M1, Canada}
\address{$^2$ Department of Physics, St. Francis Xavier University, Antigonish, Nova Scotia, B2G 2W5, Canada}
\address{$^3$ Department of Physics and Astronomy, University of Waterloo, Waterloo, Ontario, N2L 3G1, Canada}
\address{$^4$ Department of Physics, Engineering Physics, and Astronomy, Queen's University, Kingston, Ontario, K7L 3N6, Canada}
\address{$^5$ Setif 1 University-Ferhat Abbas. Faculty of Sciences. Department of Physics and Laboratory of Quantum Physics and Dynamical Systems, Setif, Algeria}

\ead{\mailto{liam.mfarrell@outlook.com}}
%\vspace{10pt}
%\begin{indented}
%\item[] February 1, 2024
%\end{indented}

\begin{abstract}
The one-body density matrix (ODM) for a zero temperature non-interacting Fermi gas can be approximately obtained in the semiclassical regime through different $\hbar$-expansion techniques.
One would expect that each method of approximating the ODM should yield equivalent density matrices which are both Hermitian and idempotent to any order in $\hbar$.
However, the Kirzhnits and Wigner-Kirkwood methods do not yield these properties, while the Grammaticos-Voros method does.
Here we show explicitly, for arbitrary $d$-dimensions through an appropriate change into symmetric coordinates, that each method is indeed identical, Hermitian, and idempotent. This change of variables resolves the inconsistencies between the various methods, showing that the non-Hermitian and non-idempotent behaviour of the Kirzhnits and Wigner-Kirkwood methods is an artifact of performing a non-symmetric truncation to the
semiclassical $\hbar$-expansions. Our work also provides the first explicit derivation of the $d$-dimensional Grammaticos-Voros ODM, originally proposed by Redjati et al (2019 \textit{J. Phys. Chem. Solids} 134 313–8) based on their $d=1,2,3,4$ expressions.
\end{abstract}

\vspace{2pc}

%\submitto{\jpa}

%\maketitle


\eqnobysec

\newpage
\section{Introduction}\label{sec:intro}

Density functional theory (DFT) is a powerful tool used to solve quantum many-body problems. {It is applied in various fields of physics at the quantum level such as the calculation of electronic structure and band structures in solid-state physics, and it can also be applied to calculate binding energies of molecules in chemistry (see for instance:~\cite{dreizlerdensity1990,jonesdensity2015,parryangdensity1994,kryachkoludena2014,glossmanmitnikdensity2022,crisostomoetal2023} and references therein). In DFT, the local particle density $\rho(\mathbf{r})$ constitutes the basic variable to describe the properties of a system of $N$ interacting particles. For a practical use of DFT, Kohn and Sham (KS) proposed to replace the real interacting system with a system of $N$ non-interacting fermions \cite{Kohn_1965,Kohn_1999}. The ground state of the latter auxiliary system is characterized at every point in space by the same particle density as that of the real system. The KS version of DFT is a one particle theory where each fermion moves in an effective one-body potential $v_{\mathrm{eff}}(\mathbf{r})$. Let $\rho(\mathbf{r})$ be the ground state particle density; in (KS) DFT the non-interacting kinetic energy functional $T_{s}[\rho]=\int\tau(\mathbf{r})d^{d}r$ constitutes an important part of the total-energy, where the superscript $d$ within the integral represents the number of dimensions of the system. Here $\tau(\mathbf{r})$ is the kinetic energy density and is valid for a system at zero absolute temperature~\cite{dreizlerdensity1990, parryangdensity1994} in terms of the (KS) single particle states $\{\phi_{j}\}$ with corresponding eigenvalues $\{\varepsilon_{j}\}$, and is given as
\begin{equation}
    \tau(\mathbf{r})=\frac{\hbar^{2}}{2m}\sum_{\varepsilon_{j}\leq\mu}\vert \nabla\phi_{j}(\mathbf{r})\vert^{2}\label{kinEng},
\end{equation}
where $\mu$ is the Fermi energy. It is possible to express the above positive defined kinetic energy density in terms of the one-body density matrix (ODM), $\rho(\mathbf{r},\mathbf{r}')$, given in terms of (KS) single particle states by, $\rho(\mathbf{r},\mathbf{r}')=\sum_{\varepsilon_{j}\leq\mu}\phi_{j}(\mathbf{r})\phi_{j}^{\ast}(\mathbf{r}')$, so that
\begin{equation}
    \tau(\mathbf{r})=\frac{\hbar^{2}}{2m}\big(\nabla_{\mathbf{r}}\cdot\nabla_{\mathbf{r}'}\big)\rho(\mathbf{r},\mathbf{r}')\vert_{\mathbf{r}'=\mathbf{r}} .  \label{kinEng2}
\end{equation}
Notice that the ODM may be seen as the matrix element in coordinate representation of the density operator, $\hat{\rho}=\sum_{\varepsilon_{j}\leq\mu}\vert\phi_{j}\rangle\langle\phi_{j}\vert$. The diagonal element $\rho(\mathbf{r},\mathbf{r}'=\mathbf{r})\equiv\rho(\mathbf{r})$ is the local density and normalized to the total particle number $N$ of fermions, $\int\rho(\mathbf{r})d^{d}r=N$.

The local density approximation (LDA) is known as the starting point of many approximations of $\rho(\mathbf{r}, \mathbf{r}')$, and yields the well-known Thomas-Fermi kinetic energy functional \cite{vanzylanalytical2003,bracksemiclassical2018}.} Although this approximation is {a good} starting point, it often fails to capture the complete physics of a given non-interacting Fermi gas since it neglects the effects of inhomogeneities. For non-uniform systems with finite size, the exact {analytical expressions of ODMs and the resulting exact density functionals} are in general not known, but density gradient corrections to the LDA exist and can be computed through various techniques ~\cite{bracksemiclassical2018,hohenberginhomogeneous1964,joubertdensity1998,vanzyl2014,trappe2016}. The exact quantum ODM defined above is Hermitian, $\rho(\mathbf{r},\mathbf{r}')=\rho^{\ast}(\mathbf{r}',\mathbf{r})$, and idempotent, $\hat{\rho}^{2}=\hat{\rho}$. {These properties are satisfied if all terms in the $\hbar$ or gradient expansion of the ODM are accounted for whenever the series converges. A truncation at a given order may lead to a loss of the Hermiticity and idempotency, and clearly this symmetry violation has consequences for quantities that are derived from it \cite{Leeuwen_2013}.}

A {particular} subset of {gradient correction} techniques includes different semiclassical $\hbar$ expansions, with three popular methods being the Kirzhnitz method (KODM)~\cite{kirzhnitsfield1967,grossgradient1981,chongreinterpretation1995,salasnichkirzhnits2007}, the Wigner-Kirkwood method (WKODM)~\cite{kirkwoodquantum1933}, and the {Grammaticos-Voros} method (GVODM)~\cite{grammaticossemiclassical1979}.
In reference~\cite{putajakirzhnits2012} the KODM was derived up to second order in $\hbar$ for arbitrary $d$ {spatial} dimensions.
Furthermore, in reference~\cite{bracksemiclassical2018} the $d$-dimensional WKODM method is discussed although the ODM is not explicitly derived, and in references~\cite{redjatisemiclassical2019,bencheikhhermitian2016,bencheikhmanifestly2016} the GVODM is derived to second order in $\hbar$ for $d=1,2,3,4$-dimensions.
Although in each case the ODMs were all derived for a zero temperature non-interacting Fermi gas in the presence of a position-dependent external potential, they can still be used to account for the effects of finite temperature and internal interactions, see references~\cite{bencheikhmanifestly2016,vanzylgradient2011}. 

{In reference \cite{putajakirzhnits2012} it was found that when truncated to second order in $\hbar$, the $d$-dimensional} KODM and WKODM methods {are found to be manifestly} non-Hermitian and non-idempotent.
In contrast, the GVODM preserves Hermiticity and idempotency regardless of whichever order in $\hbar$ the semiclassical expansion is truncated at.
These issues are commented on in references~\cite{redjatisemiclassical2019,bencheikhhermitian2016,bencheikhmanifestly2016}.
The various ODMs should all be equivalent since they must yield the same physical properties of the Fermi gas, e.g., the total energy.
Therefore it is unsatisfying that the GVODM is manifestly Hermitian and idempotent while the KODM and WKODM are not.
Additionally, a {$d$}-dimensional expression for the GVODM has not yet been explicitly derived.
Although reference~\cite{redjatisemiclassical2019} proposed such an expression based on their $d=1,2,3,4$ results, a rigorous derivation was not performed.
 
The structure of this paper is as follows: In Section~\ref{sec:KODM}, we show that because the {WKODM and} KODM {are} formulated in terms of inherently \emph{non-symmetric coordinates} $(\mathbf{r},\mathbf{r}')$ from the beginning of {{their derivations}, {they do} not display the same symmetry as the GVODM which is formulated in terms of center of mass $\mathbf{R}$ and relative $\mathbf{s}$ coordinates.
    \begin{gather}
        \mathbf{R}=(\mathbf{r}+\mathbf{r}')/2, \hspace{2cm} \mathbf{s}=\mathbf{r}-\mathbf{r}'. \label{symmCoord}
    \end{gather}
We explicitly show in {arbitrary $d$}-dimensions that, upon the proper symmetrization procedure (change of coordinates) from $(\mathbf{r},\mathbf{r}')\rightarrow(\mathbf{R},\mathbf{s})$, the {{WKODM and} KODM {{are} identical to the GVODM up to second order in $\hbar$.
Our starting point relies on the ability to express each term of the {WKODM and }KODM as a specific inverse Laplace transformation before the coordinate transformation is undertaken. We show in \ref{appendix:WKDOM} that in terms of non-symmetric coordinates, the WKODM is identical to the KODM via these inverse Laplace transformations~\cite{bracksemiclassical2018,geldartconvergence1986}, allowing us to make the necessary change of coordinates.
{For completeness, in \ref{appendix:alternate} we also provide an alternate yet equivalent derivation in $d$-dimensions of the equality between the WKODM (and therefore also KODM) and the GVODM, by making the change of coordinates to $(\mathbf{R},\mathbf{s})$ from the very beginning before any inverse Laplace transformations are evaluated.}

The symmetrization procedure for both the KODM and WKODM demonstrates that regardless of whichever method is used to derive the ODM of a zero temperature non-interacting Fermi gas, the same Hermitian and idempotent expression is obtained as one would expect.
Our results can be shown to hold when higher order terms in the ODMs are accounted for, but, for simplicity, we restricted our analysis to second order in $\hbar$.
This therefore resolves the Hermiticity and idempotency inconsistency between the various methods, showing that problems only arise when non-symmetric truncations of semiclassical expansions are taken.
Additionally, this explicitly shows that the predicted {$d$}-dimensional expression for the GVODM proposed in reference~\cite{redjatisemiclassical2019} is in fact correct, as we exactly obtain their result through our symmetrization procedure for both the KODM and WKODM methods.
To our knowledge, these issues have been overlooked in the semiclassical expansion methods utilized in DFT.

\section{\texorpdfstring{Hermitian form of the {WKODM and} KODM in {$d$}-dimensions}{Hermitian form of the KODM in d-dimensions}}\label{sec:KODM}

The {$d$}-dimensional KODM for a non-interacting Fermi gas in non-symmetric coordinates $(\mathbf{r},\mathbf{r}')$ {to second order} is given by~\cite{putajakirzhnits2012}
    \begin{gather}
        \rho_{K}(\mathbf{r},\mathbf{r}')=A_{0}^{{K}}+A_{1}^{{K}}+A_{2}^{{K}}; \label{KODM}
    \end{gather}
    \begin{align}
        A_{0}^{{K}}&=\frac{g k_{F}^{d}}{(2\pi z)^{d/2}} J_{\frac{d}{2}}(z) \label{A0term}, \\
        A_{1}^{{K}}&=-\frac{g k_{F}^{d-3}z^{2}}{4(2\pi z)^{d/2}}J_{\frac{d}{2}-1}(z)\bigg(\nabla_{\mathbf{r}}k_{F}^{2}\cdot\frac{\mathbf{s}}{s}\bigg)\label{A1term}, \\
        A_{2}^{{K}}&=\frac{g k_{F}^{d-4}z^{2}}{24(2\pi z)^{d/2}}J_{\frac{d}{2}-2}(z)\big(\nabla_{\mathbf{r}}^{2}k_{F}^{2}\big)+\frac{g k_{F}^{d-6}z^{3}}{96(2\pi z)^{d/2}}J_{\frac{d}{2}-3}(z)\big(\nabla_{\mathbf{r}}k_{F}^{2}\big)^{2}\nonumber\\
        &\quad+\frac{g k_{F}^{d-6}z^{4}}{32(2\pi z)^{d/2}}J_{\frac{d}{2}-2}(z)\bigg(\nabla_{\mathbf{r}}k_{F}^{2}\cdot\frac{\mathbf{s}}{s}\bigg)^{2}\nonumber\\
        &\quad+\frac{g k_{F}^{d-4}z^{3}}{12(2\pi z)^{d/2}}J_{\frac{d}{2}-1}(z)\bigg[\nabla_{\mathbf{r}}\bigg(\nabla_{\mathbf{r}}k_{F}^{2}\cdot\frac{\mathbf{s}}{s}\bigg)\cdot\frac{\mathbf{s}}{s}\bigg].
        \label{A2term}
    \end{align}
{ The factor $g$ is the spin degeneracy, the local Fermi wavenumber is defined by $k_{F}=k_{F}(\mathbf{r})=\frac{\sqrt{2m}}{\hbar}\big(\mu-V(\mathbf{r})\big)^{1/2}$ where we have denoted $z=k_{F} s$ with $s=\vert s\vert=\sqrt{\mathbf{s}\cdot\mathbf{s}}$, and $J_{\nu}(z)$ is the cylindrical Bessel function of order $\nu$~\cite{gradshteintable2015}. For convenience, we have suppressed the $\mathbf{r}$ coordinate dependence of both $k_{F}$ and the $A_{0,1,2}^{{K}}$ terms.}

\Cref{KODM} {with \cref{A0term,A1term,A2term}} was obtained by simplifying equation {(}17{)} from reference~\cite{putajakirzhnits2012} using the Bessel function identities described in \ref{appendix:relations}, namely \cref{Bessel1,Bessel2}.
Furthermore, we have collected the various terms of the {K}ODM based on their order of $\hbar$ (or equivalently, their order of {gradient of the potential}).
{In \cref{A0term},} $A_{0}^{{K}}$ is the classical zeroth order term of the semiclassical expansion { (order $\hbar^{0}$)}, also known as the Thomas-Fermi term.
It is the only term which does not vanish {for uniform systems. To account for effects due to inhomogeneities, one must go beyond the LDA by including density gradient corrections through the terms $A_{1}^{K}$ and $A_{2}^{K}$ which are respectively of order $\hbar$ and $\hbar^{2}$.}

\Cref{KODM} is manifestly non-Hermitian and non-idempotent as discussed in references~\cite{redjatisemiclassical2019,bencheikhhermitian2016,bencheikhmanifestly2016}. { Since \cref{KODM} contains no complex variables and therefore is equal to its own complex conjugate, the non-Hermiticity can be seen by swapping $\mathbf{r}$ and $\mathbf{r}'$, equivalent to the replacement $\mathbf{s}\rightarrow -\mathbf{s}$, and observing that the resulting expression is not equal to $\rho_{K}(\mathbf{r},\mathbf{r}')$. That is, $\rho_{K}(\mathbf{r},\mathbf{r}')\neq \rho_{K}^{\ast}(\mathbf{r}',\mathbf{r})$.}
However, by changing to center of mass $\mathbf{R}$ and relative coordinates $\mathbf{s}$ via \cref{symmCoord}, we show that \cref{KODM} can be {recasted} into a manifestly Hermitian and idempotent form identical to the GVODM.
The transformation from $(\mathbf{r},\mathbf{r}')\rightarrow (\mathbf{R},\mathbf{s})$ variables is non-trivial as we will show.
Consequently, to simplify the process, we will individually transform each {$A_{0,1,2}^{{K}}$} term in \cref{KODM}, and sum the results in the end via \cref{KODM}. 

Although {in reference~\cite{putajakirzhnits2012} \cref{KODM,A0term,A1term,A2term} are derived in $d$}-dimensions in terms of non-symmetric coordinates using the commutator method of Kirzhnitz~\cite{kirzhnitsfield1967}, they can equivalently also be derived in these coordinates via the WKODM method~\cite{bracksemiclassical2018,kirkwoodquantum1933,geldartconvergence1986}.
We show this to be true in \ref{appendix:WKDOM} so that \cref{A0term,A1term,A2term} may alternatively each be expressed {as an appropriate inverse Laplace transform of the so-called Bloch density matrix or Bloch propagator. The latter is of great interest in the study of properties
of non-interacting fermions in a one-body Hamiltonian $H$. The Bloch density matrix is defined as $C(\mathbf{r},\mathbf{r}';\beta)=\langle\mathbf{r}\vert\textrm{e}^{\beta H}\vert\mathbf{r}'\rangle$, where $\beta$ is a variable which, in general, is taken to be complex, and not the inverse of the temperature. In terms of single particle wavefunctions, the Bloch density matrix reads $C(\mathbf{r},\mathbf{r}';\beta)=\sum_{j}\phi_{j}(\mathbf{r})\phi_{j}^{\ast}(\mathbf{r}')\textrm{e}^{\beta\varepsilon_{j}}$. By expressing the ODM in terms of the unit step function $\theta(x)$ as $\rho(\mathbf{r},\mathbf{r}')=\sum_{j}\phi_{j}(\mathbf{r})\phi_{j}^{\ast}(\mathbf{r}')\theta(\mu-\varepsilon_{j})$, where the sum is extended over the whole spectrum, this latter ODM, for a given Fermi energy $\mu$, can be written as an inverse Laplace transform of $C(\mathbf{r},\mathbf{r}';\beta)/\beta$ \cite{bracksemiclassical2018}
\begin{gather}
    \rho(\mathbf{r},\mathbf{r}')=\frac{1}{2\pi i}\int_{c-i\infty}^{c+i\infty}d\beta\;\textrm{e}^{\beta\mu}\frac{C(\mathbf{r},\mathbf{r}';\beta)}{\beta}=\mathcal{L}^{-1}_{\mu}\bigg\{\frac{C(\mathbf{r},\mathbf{r}';\beta)}{\beta}\bigg\}, \label{blochProp}
\end{gather}
where we have used the following identity for $x=\mu-\varepsilon_{j}$ and $z=\beta$ \cite{HBMF_1970}: $\theta(x)=\frac{1}{2\pi i}\int_{c-i\infty}^{c+i\infty}\frac{\textrm{e}^{xz}}{z}dz=\mathcal{L}_{x}^{-1}\{z^{-1}\}$, and where $c$ is a real number greater than zero. 

Now returning to the semiclassical expansion considered in this work, an expansion in terms of powers of $\hbar$ is known for $C(\mathbf{r},\mathbf{r}';\beta)$, and is called the Wigner-Kirkwood expansion \cite{bracksemiclassical2018}. Up to second order we write it as $C^{WK}(\mathbf{r},\mathbf{r}';\beta)\approx C^{WK}_{0}(\mathbf{r},\mathbf{r}';\beta)+C^{WK}_{1}(\mathbf{r},\mathbf{r}';\beta)+C^{WK}_{2}(\mathbf{r},\mathbf{r}';\beta)$ where the first term is the zeroth order term (Thomas-Fermi), the second term is of order $\hbar$, and the third term is of order $\hbar^{2}$. Substituting the latter $\hbar$-expansion of $C(\mathbf{r},\mathbf{r}';\beta)$ into \cref{blochProp} allows one to obtain the ODM in the Wigner-Kirkwood approach in the form $\rho_{WK}(\mathbf{r},\mathbf{r}')=A_{0}^{WK}+A_{1}^{WK}+A_{2}^{WK}$ with}
    \begin{gather}
       A_{0}^{{WK}}=\mathcal{L}^{-1}_{\mu}\bigg\{\frac{C_{0}^{{WK}}(\mathbf{r},\mathbf{r}';\beta)}{\beta}\bigg\} \label{A0term2}, \\
        A_{1}^{{WK}}=\mathcal{L}^{-1}_{\mu}\bigg\{\frac{C_{1}^{{WK}}(\mathbf{r},\mathbf{r}';\beta)}{\beta}\bigg\} \label{A1term2}, \\
        A_{2}^{{WK}}=\mathcal{L}^{-1}_{\mu}\bigg\{\frac{C_{2}^{{WK}}(\mathbf{r},\mathbf{r}';\beta)}{\beta}\bigg\} \label{A2term2}. 
    \end{gather} 
{In the above the $C_{0,1,2}^{WK}(\mathbf{r},\mathbf{r}';\beta)$} are {respectively} given by \cite{bracksemiclassical2018}
    \begin{align}
        C_{0}^{{WK}}(\mathbf{r},\mathbf{r}';\beta)&=\frac{g}{(2\pi\hbar)^{d}}\int d^{d}p\;\textrm{e}^{-\beta H_{cl}(\mathbf{r},\mathbf{p})+\frac{i}{\hbar}\mathbf{p}\cdot\mathbf{s}} \label{Bloch0term},\\
        C_{1}^{{WK}}(\mathbf{r},\mathbf{r}';\beta)&=-\frac{i g\hbar\beta^{2}}{2m(2\pi\hbar)^{d}}\int d^{d}p\;\textrm{e}^{-\beta H_{cl}(\mathbf{r},\mathbf{p})+\frac{i}{\hbar}\mathbf{p}\cdot\mathbf{s}}\Big(\nabla_{\mathbf{r}}V(\mathbf{r})\cdot\mathbf{p}\Big)  \label{Bloch1term},\\
        C_{2}^{{WK}}(\mathbf{r},\mathbf{r}';\beta)&=\frac{g\hbar^{2}}{(2\pi\hbar)^{d}}\int d^{d}p\;\textrm{e}^{-\beta H_{cl}(\mathbf{r},\mathbf{p})+\frac{i}{\hbar}\mathbf{p}\cdot\mathbf{s}}\bigg(-\frac{\beta^{{2}}}{4m}\nabla_{\mathbf{r}}^{2}V(\mathbf{r}) +\frac{\beta^{3}}{6m}\big(\nabla_{\mathbf{r}}V(\mathbf{r})\big)^{2}\nonumber \\
        &\hspace{3.5cm}-\frac{\beta^{4}}{8m^{2}}\big({\nabla_{\mathbf{r}}V(\mathbf{r})\cdot\mathbf{p}}\big)^{2}+\frac{\beta^{3}}{6m^{2}}(\mathbf{p}\cdot\nabla_{\mathbf{r}})^{2}V(\mathbf{r})\bigg) \label{Bloch2term}.
    \end{align}
Here $H_{cl}(\mathbf{r},\mathbf{p})=\frac{\mathbf{p}^{2}}{2m}+V(\mathbf{r})$ is the classical counterpart of the quantum one-body Hamiltonian.
Substituting \cref{Bloch0term,Bloch1term,Bloch2term} into \cref{A0term2,A1term2,A2term2} and applying the inverse Laplace transform relationship provided in \ref{appendix:relations} by \cref{invLap1}{, we prove in \ref{appendix:WKDOM} that this} exactly yields \cref{A0term,A1term,A2term}{, i.e., we find $A_{0}^{K}=A_{0}^{WK}$, $A_{1}^{K}=A_{1}^{WK}$, and $A_{2}^{K}=A_{2}^{WK}$. It then follows that the ODM of Kirzhnits and Wigner-Kirkwood are found to be identical, $\rho_{K}(\mathbf{r},\mathbf{r}')=\rho_{WK}(\mathbf{r},\mathbf{r}')$. Notice that this equality holds in $d$-dimensions and in the derivation we used the $d$-dimensional Gaussian integral in \cref{Bloch0term} for the momentum $\mathbf{p}$ variable, which straightforwardly yields
    \begin{gather}
        C^{WK}_{0}(\mathbf{r},\mathbf{r}';\beta)=g\bigg(\frac{m}{2\pi\beta\hbar^{2}}\bigg)^{d/2}\textrm{e}^{-\frac{m \mathbf{s}^{2}}{2\hbar^{2}\beta}-\beta V(\mathbf{r})}  \label{Bloch0termB}.
    \end{gather}}
{The above analytical expression is very important since it} offers{, as we will show,} the starting point for our symmetrization procedure{. It turns out to be more convenient to first} transform the {spatial} variables {in} \cref{Bloch0term,Bloch1term,Bloch2term} from $(\mathbf{r},\mathbf{r}')\rightarrow(\mathbf{R},\mathbf{s})$, and after that perform the inverse Laplace transforms in \cref{A0term2,A1term2,A2term2} {to finally get the ODM in terms of center of mass and relative variables. Our final goal is to compare the resulting expression denoted now by $\rho_{WK}(\mathbf{R},\mathbf{s})$, with the ODM derived in reference~\cite{redjatisemiclassical2019} via the Grammaticos-Voros expansion, which is directly expressed in terms of $\mathbf{R}$ and $\mathbf{s}$ variables.

For the sake of pedagogical clarity, we will perform the transformation of variables} to each term separately, starting with $A_{0}^{{WK}}$ and ending with $A_{2}^{{WK}}$ {(we recall that these terms are identical to the $A_{0,1,2}^{{K}}$ of the Kirzhnitz ODM)}. The results will be summed afterwards via \cref{KODM} to obtain {equivalently the WKODM or} the KODM in terms of the symmetric coordinates $(\mathbf{R},\mathbf{s})$.

\subsection{\texorpdfstring{{Change of variables $(\mathbf{r},\mathbf{r}')\rightarrow(\mathbf{R},\mathbf{s})$ in $A_{0}^{WK}$}}{A0 change of variables}}

{In order to evaluate the inverse Laplace transform of \cref{A0term2} in symmetric coordinates, we first must transform \cref{Bloch0termB} via \cref{symmCoord}. We consider the Taylor expansion of the potential $V(\mathbf{r})$ around the non-locality $s$, and up to second order derivatives of the potential we have}
    \begin{gather}
        V(\mathbf{r})= V(\mathbf{R}+\mathbf{s}/2)\approx V(\mathbf{R})+\frac{1}{2}\big(\nabla_{\mathbf{R}}V(\mathbf{R})\cdot\mathbf{s}\big)+\frac{1}{8}\Big[\nabla_{\mathbf{R}}\big(\nabla_{\mathbf{R}}V(\mathbf{R})\cdot\mathbf{s}\big)\cdot\mathbf{s}\Big]. \label{Vseries}
    \end{gather}
{Now, using the Taylor expansion in \cref{Vseries}, we can write
\begin{gather}
    \textrm{e}^{-\beta V(\mathbf{r})}\approx\textrm{e}^{-\beta V(\mathbf{R})}\textrm{e}^{-\beta\bigg(\frac{1}{2}\big(\nabla_{\mathbf{R}}V(\mathbf{R})\cdot\mathbf{s}\big)+\frac{1}{8}\Big[\nabla_{\mathbf{R}}\big(\nabla_{\mathbf{R}}V(\mathbf{R})\cdot\mathbf{s}\big)\cdot\mathbf{s}\Big]\bigg)} \label{tExpan1}.
\end{gather}
Expanding the second exponential in the right side of \cref{tExpan1} and retaining only terms up to the second derivative of the potential $V(\mathbf{R})$ we find
\begin{align}
    \textrm{e}^{-\beta V(\mathbf{r})}\approx \textrm{e}^{-\beta V(\mathbf{R})}&\Bigg[1-\beta\bigg(\frac{1}{2}\big(\nabla_{\mathbf{R}}V(\mathbf{R})\cdot\mathbf{s}\big)+\frac{1}{8}\Big[\nabla_{\mathbf{R}}\big(\nabla_{\mathbf{R}}V(\mathbf{R})\cdot\mathbf{s}\big)\cdot\mathbf{s}\Big]\bigg) \nonumber\\
    &+\frac{\beta^{2}}{8}\big(\nabla_{\mathbf{R}}V(\mathbf{R})\cdot\mathbf{s}\big)^{2}\Bigg] \label{tExpan2}.
\end{align}
Substituting \cref{tExpan2} into \cref{Bloch0termB} then allows us to write \cref{A0term} as}
    \begin{gather}
        A_{0}^{{WK}}\approx X_{a}+X_{b}+X_{c}; \label{A0term3} 
    \end{gather}
    \begin{align}
        X_{a}&=g \bigg(\frac{m}{2\pi\hbar^{2}}\bigg)^{d/2}\mathcal{L}_{\mu}^{-1}\bigg\{\beta^{-d/2-1}\textrm{e}^{-\frac{m\mathbf{s}^{2}}{2\hbar^{2}\beta}-\beta V(\mathbf{R})}\bigg\} \label{X1term}, \\
        X_{b}&=
        -g\bigg(\frac{m}{2\pi\hbar^{2}}\bigg)^{d/2}\bigg(\frac{1}{2}\big(\nabla_{\mathbf{R}}V(\mathbf{R})\cdot\mathbf{s}\big)+\frac{1}{8}\Big[\nabla_{\mathbf{R}}\big(\nabla_{\mathbf{R}}V(\mathbf{R})\cdot\mathbf{s}\big)\cdot\mathbf{s}\Big]\bigg)\nonumber\\
        &\quad\times\mathcal{L}_{\mu}^{-1}\bigg\{\beta^{-d/2}\textrm{e}^{-\frac{m\mathbf{s}^{2}}{2\hbar^{2}\beta}-\beta V(\mathbf{R})}\bigg\},\label{X2term}\\
        X_{c}&=\frac{g}{8}\bigg(\frac{m}{2\pi\hbar^{2}}\bigg)^{d/2}\big(\nabla_{\mathbf{R}}V(\mathbf{R})\cdot\mathbf{s}\big)^{2}\mathcal{L}_{\mu}^{-1}\bigg\{\beta^{-d/2+1}\textrm{e}^{-\frac{m\mathbf{s}^{2}}{2\hbar^{2}\beta}-\beta V(\mathbf{R})}\bigg\}, \label{X3term} 
    \end{align}
Each inverse Laplace transform in \cref{X1term,X2term,X3term} can be evaluated via \cref{invLap1} which is derived in \ref{appendix:relations}. 
After substituting the result and simplifying, the expression for $A_{0}^{{WK}}$ in symmetric $(\mathbf{R},\mathbf{s})$ coordinates is obtained and given by
    \begin{align}
        A_{0}^{{WK}}&=\frac{g k_{F}^{d}}{(2\pi z)^{d/2}}J_{\frac{d}{2}}(z)+\frac{g k_{F}^{d-3}z^{2}}{4(2\pi z)^{d/2}}J_{\frac{d}{2}-1}(z)\bigg(\nabla_{\mathbf{R}}k_{F}^{2}\cdot\frac{\mathbf{s}}{s}\bigg)\nonumber\\
        &\quad+\frac{g k_{F}^{d-4}z^{3}}{16(2\pi z)^{d/2}}J_{\frac{d}{2}-1}(z)\bigg[\nabla_{\mathbf{R}}\bigg(\nabla_{\mathbf{R}}k_{F}^{2}\cdot\frac{\mathbf{s}}{s}\bigg)\cdot\frac{\mathbf{s}}{s}\bigg]\nonumber\\
        &\quad+\frac{g k_{F}^{d-6}z^{4}}{32(2\pi z)^{d/2}}J_{\frac{d}{2}-2}(z)\bigg(\nabla_{\mathbf{R}}k_{F}^{2}\cdot\frac{\mathbf{s}}{s}\bigg)^{2}.\label{A0term4}
   \end{align}
We also used the definition $k_{F}=k_{F}(\mathbf{R})=\frac{\sqrt{2m}}{\hbar}\big(\mu-V(\mathbf{R})\big)^{1/2}$ so that $k_{F}$ is now a function of $\mathbf{R}$. This will also be the case for the symmetrized $A_{1}^{{WK}}$ and $A_{2}^{{WK}}$ terms which will be derived later.

\subsection{\texorpdfstring{{Change of variables $(\mathbf{r},\mathbf{r}')\rightarrow(\mathbf{R},\mathbf{s})$ in $A_{1}^{{WK}}$}}{A1 change of variables}}

Before changing the coordinates of \cref{Bloch1term}, it can be re-expressed  {using the identity $\nabla_{\mathbf{s}}\textrm{e}^{i\mathbf{p}\cdot\mathbf{s}/\hbar}=\frac{i\mathbf{p}}{\hbar}\textrm{e}^{i\mathbf{p}\cdot\mathbf{s}/\hbar}$}, and {by} solving the resulting {$d$}-dimensional Gaussian integral in a similar manner to the previous {sub}section. This gives
     \begin{align}
        C_{1}(\mathbf{r},\mathbf{r}';\beta)&=-\frac{i g\hbar\beta^{2}}{2m(2\pi\hbar)^{d}}\int d^{d}p\;\textrm{e}^{-\beta H_{cl}(\mathbf{r},\mathbf{p})+\frac{i}{\hbar}\mathbf{p}\cdot\mathbf{s}}\Big(\nabla_{\mathbf{r}}V(\mathbf{r})\cdot\mathbf{p}\Big)\nonumber\\   
        &=-\frac{g\hbar^{2}\beta^{2}}{2m(2\pi\hbar)^{d}}\Big(\nabla_{\mathbf{r}}V(\mathbf{r})\cdot\nabla_{\mathbf{s}}\Big)\int d^{d}p\;\textrm{e}^{-\beta H_{cl}(\mathbf{r},\mathbf{p})+\frac{i}{\hbar}\mathbf{p}\cdot\mathbf{s}}\nonumber\\ 
        &=-\frac{g m^{d/2-1}\hbar^{2-d}}{2(2\pi)^{d/2}}\Big(\nabla_{\mathbf{r}}V(\mathbf{r})\cdot\nabla_{\mathbf{s}}\Big)\beta^{-d/2+2}\textrm{e}^{-\frac{m\mathbf{s}^{2}}{2\hbar^{2}\beta}-\beta V(\mathbf{r})}\label{Bloch1termB}.
    \end{align}
For symmetrization, we {first} make use of the approximate gradient identity derived in \ref{appendix:relations}, \cref{gradr1}, to transform $\nabla_{\mathbf{r}}V(\mathbf{r})$ into $(\mathbf{R},\mathbf{s})$ coordinates. Combining this with \cref{Bloch1termB,A1term2}, {while making use of the Taylor expansion \cref{tExpan2} in a similar way to how it was previously used for \cref{A0term3},} yields
    \begin{gather}
       A_{1}^{{WK}}\approx Y_{a}+Y_{b}+Y_{c}; \label{A1term3}
    \end{gather}
    \begin{align}
       Y_{a}&=-\frac{g m^{d/2-1}\hbar^{2-d}}{2(2\pi)^{d/2}}\big(\nabla_{\mathbf{R}}V(\mathbf{R})\cdot\nabla_{\mathbf{s}}\big)\mathcal{L}_{\mu}^{-1}\bigg\{\beta^{-d/2+1}\textrm{e}^{-\frac{m\mathbf{s}^{2}}{2\hbar^{2}\beta}-\beta V(\mathbf{R})}\bigg\} \label{Y1term}, \\ 
       Y_{b}&=\frac{g m^{d/2-1}\hbar^{2-d}}{4(2\pi)^{d/2}}\big(\nabla_{\mathbf{R}}V(\mathbf{R})\cdot\mathbf{s}\big)\big(\nabla_{\mathbf{R}}V(\mathbf{R})\cdot\nabla_{\mathbf{s}}\big)\mathcal{L}_{\mu}^{-1}\bigg\{\beta^{-d/2+2}\textrm{e}^{-\frac{m\mathbf{s}^{2}}{2\hbar^{2}\beta}-\beta V(\mathbf{R})}\bigg\} \label{Y2term}, \\
       Y_{c}&=-\frac{g m^{d/2}\hbar^{2-d}}{4(2\pi)^{d/2}}\Big[\nabla_{\mathbf{R}}\big(\nabla_{\mathbf{R}}V(\mathbf{R})\cdot\mathbf{s}\big)\cdot\nabla_{\mathbf{s}}\Big]\mathcal{L}_{\mu}^{-1}\bigg\{\beta^{-d/2+1}\textrm{e}^{-\frac{m\mathbf{s}^{2}}{2\hbar^{2}\beta}-\beta V(\mathbf{R})}\bigg\}\label{Y3term}.
    \end{align}
Here we have not separated each term based on {its} power of $\beta$ like we did for \cref{X1term,X2term,X3term}, and instead separate them based on $\nabla_{\mathbf{R}}$.
Using the inverse Laplace transformation identity {of }\cref{invLap1} as before{, along with} the gradient and Bessel function relationships of \cref{gradS,Bessel3} from \ref{appendix:relations}, we can evaluate and simplify \cref{A1term3,Y2term,Y3term} to find $A_{1}^{{WK}}$ in terms of symmetric coordinates
    \begin{align}
        A_{1}^{{WK}}&=-\frac{g k_{F}^{d-3}z^{2}}{4(2
        \pi z)^{d/2}}J_{\frac{d}{2}-1}(z)\bigg(\nabla_{\mathbf{R}}k_{F}^{2}\cdot\frac{\mathbf{s}}{s}\bigg)-\frac{g k_{F}^{d-6}z^{4}}{16(2\pi z)^{d/2}}J_{\frac{d}{2}-2}(z)\bigg(\nabla_{\mathbf{R}}k_{F}^2\cdot\frac{\mathbf{s}}{s}\bigg)^{2} \nonumber \\
        &\quad-\frac{g k_{F}^{d-4}z^{3}}{8(2\pi z)^{d/2}}J_{\frac{d}{2}-1}(z)\bigg[\nabla_{\mathbf{R}}\bigg(\nabla_{\mathbf{R}}k_{F}^{2}\cdot\frac{\mathbf{s}}{s}\bigg)\cdot\frac{\mathbf{s}}{s}\bigg] \label{A1term4}.
    \end{align}

\subsection{\texorpdfstring{{Change of variables $(\mathbf{r},\mathbf{r}')\rightarrow(\mathbf{R},\mathbf{s})$ in $A_{2}^{{WK}}$}}{A2 change of variables}}

To symmetrize the final term $A_{2}^{{WK}}$, we take a different approach which avoids the need to perform the coordinate transformation before the inverse Laplace transform is evaluated.
Since we are only keeping up to second order terms, and every term within the integrand of \cref{Bloch2term} is already second order in $\nabla_{\mathbf{r}}$, when expanding the {$\textrm{e}^{-\beta V(\mathbf{r})}$ factor in \cref{Bloch2term}, as shown by \cref{tExpan2}}, only the zeroth order part ${\textrm{e}^{-\beta V(\mathbf{R})}}$ is retained.
Therefore, no additional second order or lower terms appear in the symmetric expression for $A_{2}^{{WK}}$.
This is in contrast to $A_{0}^{{WK}}$ and $A_{1}^{{WK}}$ where the additional terms $X_{b,c}$ and $Y_{b,c}$ appeared, respectively.
Thus for $A_{2}^{{WK}}$, we can simply directly apply our transformation of variables \cref{symmCoord} to the original non-symmetric KODM expression of $A_{2}^{{K}}$ {(equivalent to the non-symmetric WKODM)}, \cref{A2term}.
To do this, we use three gradient identities given by \cref{gradr1,gradr2,gradr3} as shown and derived in \ref{appendix:relations}. Substituting the identities into \cref{A2term} yields the symmetric value of $A_{2}^{{WK}}$, which after simplification is given by
    \begin{align}
        A_{2}^{{WK}}&\approx\frac{g k_{F}^{d-4}z^{2}}{24(2\pi z)^{d/2}}J_{\frac{d}{2}-2}(z)\big(\nabla_{\mathbf{R}}^{2}k_{F}^{2}\big)+\frac{g k_{F}^{d-6}z^{3}}{96(2\pi z)^{d/2}}J_{\frac{d}{2}-3}(z)\big(\nabla_{\mathbf{R}}k_{F}^{2}\big)^{2}\nonumber\\
        &\quad+\frac{g k_{F}^{d-6}z^{4}}{32(2\pi z)^{d/2}}J_{\frac{d}{2}-2}(z)\bigg(\nabla_{\mathbf{R}}k_{F}^{2}\cdot\frac{\mathbf{s}}{s}\bigg)^{2}\nonumber\\
        &\quad+\frac{g k_{F}^{d-4}z^{3}}{12(2\pi z)^{d/2}}J_{\frac{d}{2}-1}(z)\bigg[\nabla_{\mathbf{R}}\bigg(\nabla_{\mathbf{R}}k_{F}^{2}\cdot\frac{\mathbf{s}}{s}\bigg)\cdot\frac{\mathbf{s}}{s}\bigg].
       \label{A2term3} 
    \end{align}

\subsection{\texorpdfstring{Symmetrized {WKODM and} KODM}{Symmetrized WKODM and KODM}}

We are finally in a position to obtain {both} the symmetric {WKODM and} KODM by combining the symmetrized {$A_{0,1,2}^{{WK}}$} terms. {Recall that the WKODM is given as $\rho_{WK}(\mathbf{R},\mathbf{s})=A_{0}^{WK}+A_{1}^{WK}+A_{2}^{WK}$, and using the results given respectively in }\cref{A0term4,A1term4,A2term3} into \cref{KODM}, after the collection and cancellation of identical terms, we find the final result 
    \begin{empheq}[box=\fbox]{align}
         {\rho_{WK}(\mathbf{R},\mathbf{s})}=\rho_{K}(\mathbf{R},\mathbf{s})&=\frac{g k_{F}^{d}}{(2\pi z)^{d/2}}J_{\frac{d}{2}}(z)+\frac{g k_{F}^{d-4}z^{2}}{24(2\pi z)^{d/2}}J_{\frac{d}{2}-2}(z)\big(\nabla_{\mathbf{R}}^{2}k_{F}\big) \nonumber \\
        &\quad+\frac{g k_{F}^{d-6}z^{3}}{96(2\pi z)^{d/2}}J_{\frac{d}{2}-3}(z)\big(\nabla_{\mathbf{R}}k_{F}\big)^{2}\nonumber\\
        &\quad+\frac{g k_{F}^{d-4}z^{3}}{48 (2\pi z)^{d/2}}{J_{\frac{d}{2}-1}(z)}\bigg[\nabla_{\mathbf{R}}\bigg(\nabla_{\mathbf{R}}k_{F}\cdot\frac{\mathbf{s}}{s}\bigg)\cdot\frac{\mathbf{s}}{s}\bigg].\label{GVODM}
    \end{empheq}
{The above expression constitutes the key result of this paper, and} is exactly identical to the {$d$}-dimensional GVODM originally proposed by reference~\cite{redjatisemiclassical2019} based upon their $d=1,2,3,4$ derivations{, see equation (4.13) in the aforementioned reference}.
Here we have provided an explicit derivation of their result starting from the known {$d$}-dimensional KODM expression of reference~\cite{putajakirzhnits2012} in non-symmetric coordinates (equivalent to the WKODM), and have shown that upon the proper change to symmetric coordinates, the WKODM, KODM, and GVODM are identical, ${\rho_{WK}(\mathbf{R},\mathbf{s})=}\rho_{K}(\mathbf{R},\mathbf{s})=\rho_{GV}(\mathbf{R},\mathbf{s})$.
The losses of both Hermiticity and idempotency of the {Wigner-Kirkwood and Kirzhnits ODMs} are only an artifact of non-symmetric truncations of the respective semiclassical $\hbar$-expansions.

\section{Concluding remarks}\label{sec:summary}

In summary we have studied the approximate semiclassical ODM of a zero temperature non-interacting Fermi gas obtained through three different gradient expansion techniques: the KODM, WKODM, and GVODM methods.
We have resolved the inconsistencies discussed in the literature~\cite{redjatisemiclassical2019,bencheikhhermitian2016,bencheikhmanifestly2016}, and have shown that, regardless of whichever method is used, the ODMs are Hermitian, idempotent, and identical when cast into symmetric coordinates $(\mathbf{R},\mathbf{s})$.
Inconsistencies only arise when the semiclassical expansions are truncated in terms of non-symmetric coordinates $(\mathbf{r},\mathbf{r}')$, which occur with the non-symmetric KODM [\cref{KODM}] and WKODM methods. Additionally, we have provided {an} explicit derivation of the {$d$}-dimensional GVODM originally proposed by reference~\cite{redjatisemiclassical2019} based on their $d=1,2,3,4$ expressions. This was accomplished in Section~\ref{sec:KODM} by showing that the KODM and WKODM are equivalent in non-symmetric coordinates, and proceeding to transform the known {$d$}-dimensional expression of the {W}KODM into symmetric coordinates. {For completeness, in \ref{appendix:alternate} we also provided an alternate yet equivalent derivation, starting from the definition of the Wigner-Kirkwood Bloch density matrix.}
For simplicity, we only dealt with semiclassical terms up to second order in $\hbar$, nevertheless, our results still hold when higher order terms in the ODMs are accounted for.

\ack
L.M.F., D.E., and P.C. dedicate this paper to the memory of their late supervisor, Brandon van Zyl, who passed away in 2020.
This work originally began as an undergraduate honours thesis research project between the authors during the summer of 2017.
The goal was to follow up on comments in reference~\cite{bencheikhmanifestly2016}, and address the subtle and consistently overlooked gap in the literature (see references/comments 11 and 12 in reference~\cite{bencheikhmanifestly2016}).
We decided to continue where we left off, and finally completed this work to publish it in honour of B.V.Z., which was the original intention before his untimely passing.
This work would not have been possible without the contribution of B.V.Z.
The authors would like to thank the Natural Sciences and Engineering Research Council of Canada (NSERC) for funding during the initial stages of this work in 2017.
We would also like to thank Karl-Peter Marzlin, Duncan O'Dell, and Wyatt Kirkby for useful discussions.

\section*{Author Contributions Statement}
B.V.Z. and K.B. conceived the research, and performed the initial calculations. L.M.F., P.C., and D.E. completed the calculations, and wrote the manuscript.

\section*{Competing Interests Statement}
The authors have no competing interests to declare.

\appendix
\section{{Equivalence of KODM and} WKODM}\label{appendix:WKDOM}

{The aim of this appendix is to apply the Wigner-Kirkwood expansion of the Bloch density matrix to derive, in non-symmetric coordinates, an expression of the WKODM and to prove the equivalence of the WKODM and KODM in $d$-dimensions, $\rho_{WK}(\mathbf{r},\mathbf{r}')=\rho_{K}(\mathbf{r},\mathbf{r}')$, given explicitly by \cref{KODM}.}

In the non-symmetric {coordinates the $d$}-dimensional Bloch matrix is approximated to second order in $\hbar$ by $C^{WK}(\mathbf{r},\mathbf{r}';\beta)\approx C_{0}^{WK}(\mathbf{r},\mathbf{r}';\beta)+C_{1}^{WK}(\mathbf{r},\mathbf{r}';\beta)+C_{2}^{WK}(\mathbf{r},\mathbf{r}';\beta)${, and as shown in Section~\ref{sec:KODM} the resulting WKODM is written as $\rho_{WK}(\mathbf{r},\mathbf{r}')= A_{0}^{WK}+A_{1}^{WK}+A_{2}^{WK}$, where the $A_{0,1,2}^{WK}$ are respectively given \cref{A0term2,A1term2,A2term2} together with \cref{Bloch0term,Bloch1term,Bloch2term}. We now proceed with the evaluation of each $A_{0,1,2}^{WK}$ term. Using \cref{A0term2} with \cref{Bloch0term} we have} 
    \begin{gather*}   
    A_{0}^{WK}=g \bigg(\frac{m}{2\pi\hbar^{2}}\bigg)^{d/2}\mathcal{L}_{\mu}^{-1}\bigg\{\beta^{-d/2-1}\textrm{e}^{-\frac{m\mathbf{s}^{2}}{2\hbar^{2}\beta}-\beta V(\mathbf{r})}\bigg\},
    \end{gather*}
{and making use of the identity of \cref{invLap1} for $\xi=1$, we can write}
    \begin{gather*}
        A_{0}^{WK}= \frac{g k_{F}^{d}}{(2\pi z)^{d/2}}J_{\frac{d}{2}}(z),
    \end{gather*}
{with $z=k_{F}s$ and $k_{F}=\frac{\sqrt{2m}}{\hbar}(\mu-V(\mathbf{r}))^{1/2}$. This above result is identical to that given in \cref{A0term} for the first term $A_{0}^{K}$ of the KODM, so that
    \begin{gather*}
        A_{0}^{WK}=A_{0}^{K}.
    \end{gather*}
Let us now come to the second term $A_{1}^{WK}$, given by \cref{A1term2} with \cref{Bloch1term}. Using the identity $\nabla_{\mathbf{s}}\textrm{e}^{i\mathbf{p}\cdot\mathbf{s}/\hbar}=\frac{i\mathbf{p}}{\hbar}\textrm{e}^{i\mathbf{p}\cdot\mathbf{s}/\hbar}$, we can transform \cref{Bloch1term} as follows 
    \begin{align*}
        C_{1}^{WK}(\mathbf{r},\mathbf{r}';\beta)&=-\frac{g\hbar^2\beta^2}{2m(2\pi\hbar)^d}\Big(\nabla_{\mathbf{r}}V(\mathbf{r})\cdot\nabla_{\mathbf{s}}\Big)\int d^dp\textrm{e}^{-\beta H_{cl}(\mathbf{r},\mathbf{p})+\frac{i}{\hbar}\mathbf{p}\cdot\mathbf{s}}\\
        &=-\frac{\hbar^{2}\beta^{2}}{2m}\Big(\nabla_{\mathbf{r}}V(\mathbf{r})\cdot\nabla_{\mathbf{s}}\Big)C_{0}^{WK}(\mathbf{r},\mathbf{r}';\beta) \\
        &=-\frac{g\hbar^{2}\beta^{2}}{2m}\bigg(\frac{m}{2\pi\beta\hbar^{2}}\bigg)^{d/2}\Big(\nabla_{\mathbf{r}}V(\mathbf{r})\cdot\nabla_{\mathbf{s}}\Big)\textrm{e}^{-\frac{m\mathbf{s}^{2}}{2\hbar^{2}\beta}-\beta V(\mathbf{r})} \\
        &=\frac{g\beta}{2}\bigg(\frac{m}{2\pi\beta\hbar^{2}}\bigg)^{d/2}\big(\nabla_{\mathbf{r}}V(\mathbf{r})\cdot\mathbf{s}\big)\textrm{e}^{-\frac{m\mathbf{s}^{2}}{2\hbar^{2}\beta}-\beta V(\mathbf{r})}, 
    \end{align*}
where we have successively used \cref{Bloch0term} and \cref{Bloch0termB}. Combining \cref{A1term2} with the above result, we can write}
    \begin{gather*}   
    A_{1}^{WK}={\frac{g}{2}\bigg(\frac{m}{2\pi\hbar^{2}}\bigg)^{d/2}} \big(\nabla_{\mathbf{r}}V(\mathbf{r})\cdot{\mathbf{s}}\big)\mathcal{L}_{\mu}^{-1}\bigg\{\beta^{{-d/2}}\textrm{e}^{-\frac{m\mathbf{s}^{2}}{2\hbar^{2}\beta}-\beta V(\mathbf{r})}\bigg\}, 
    \end{gather*}
{applying once again the identity of \cref{invLap1}, this time for $\xi=0$, we find
    \begin{gather*}   
    A_{1}^{WK}= \frac{g}{2}\bigg(\frac{m}{2\pi\hbar^{2}}\bigg)^{d/2}\bigg(\frac{\hbar^{2}k_{F}}{m s}\bigg)^{d/2-1}\big(\nabla_{\mathbf{r}}V(\mathbf{r})\cdot\mathbf{s}\big)J_{\frac{d}{2}-1}(z). 
    \end{gather*}
By using the relation $\nabla_{\mathbf{r}}V(\mathbf{r})=-\frac{\hbar^{2}}{2m}\nabla_{\mathbf{r}}k_{F}^{2}$ and recalling that $z=k_{F}s$, we end with}
   \begin{gather*}   
    A_{1}^{{WK}}= -\frac{g k_{F}^{d-3}z^{2}}{4(2
        \pi z)^{d/2}}J_{\frac{d}{2}-1}(z)\bigg(\nabla_{\mathbf{r}}k_{F}^{2}\cdot\frac{\mathbf{s}}{s}\bigg). 
    \end{gather*}
{This expression is identical to that given in \cref{A1term} for the second term $A_{1}^{K}$ of the KODM, thus
\begin{gather*}
    A_{1}^{WK}=A_{1}^{K}.
\end{gather*}
In a similar fashion we now calculate $A_{2}^{WK}$ of \cref{A1term2}. We first rewrite \cref{Bloch2term} as
\begin{align*}
    C_{2}^{WK}(\mathbf{r},\mathbf{r}';\beta)
    =&\bigg(-\frac{\beta^{2}}{4m}\big(\nabla_{\mathbf{r}}^{2}V(\mathbf{r})\big)+\frac{\beta^{3}}{6m}\big(\nabla_{\mathbf{r}}V(\mathbf{r})\big)^{2}\bigg)C_{0}^{WK}(\mathbf{r},\mathbf{r}';\beta) \\
    &+\frac{\hbar^{2}\beta^{4}}{8m^{2}}\Big(\nabla_{\mathbf{r}}V(\mathbf{r})\cdot\nabla_{\mathbf{s}}\Big)^{2}C_{0}^{WK}(\mathbf{r},\mathbf{r}';\beta)\\
    &-\frac{\hbar^{4}\beta^{3}}{8m^{2}}\Big(\nabla_{\mathbf{r}}\big(\nabla_{\mathbf{r}}V(\mathbf{r})\cdot\nabla_{\mathbf{s}}\big)\cdot\nabla_{\mathbf{s}}\Big)C_{0}^{WK}(\mathbf{r},\mathbf{r}';\beta).
\end{align*}
We substitute the above expression into \cref{A1term2} together with \cref{Bloch2term} and obtain}
    \begin{align*}   
    A_{2}^{WK}=&-{\frac{g\hbar^{2}}{12m}\bigg(\frac{m}{2\pi\hbar}\bigg)^{d/2}}\big(\nabla_{\mathbf{r}}^{2}V(\mathbf{r})\big)\mathcal{L}_{\mu}^{-1}\bigg\{\beta^{-d/2+1}\textrm{e}^{-\frac{m\mathbf{s}^{2}}{2\hbar^{2}\beta}-\beta V(\mathbf{r})}\bigg\} \\
    &+{\frac{g\hbar^{2}}{24m}\bigg(\frac{m}{2\pi\hbar}\bigg)^{d/2}}\big(\nabla_{\mathbf{r}}V(\mathbf{r})\big)^{2}\mathcal{L}_{\mu}^{-1}\bigg\{\beta^{-d/2+2}\textrm{e}^{-\frac{m\mathbf{s}^{2}}{2\hbar^{2}\beta} -\beta V(\mathbf{r})}\bigg\} \\
    &+{\frac{g}{8}\bigg(\frac{m}{2\pi\hbar}\bigg)^{d/2}}\big(\nabla_{\mathbf{r}}V(\mathbf{r})\cdot{\mathbf{s}}\big)^{2}\mathcal{L}_{\mu}^{-1}\bigg\{\beta^{-d/2+1}\textrm{e}^{-\frac{m\mathbf{s}^{2}}{2\hbar^{2}\beta}-\beta V(\mathbf{r})}\bigg\} \\
    &-{\frac{g}{6}\bigg(\frac{m}{2\pi\hbar}\bigg)^{d/2}}\Big(\nabla_{\mathbf{r}}\big(\nabla_{\mathbf{r}}V(\mathbf{r})\cdot{\mathbf{s}}\big)\cdot{\mathbf{s}}\Big)\mathcal{L}_{\mu}^{-1}\bigg\{\beta^{-d/2}\textrm{e}^{-\frac{m\mathbf{s}^{2}}{2\hbar^{2}\beta}-\beta V(\mathbf{r})}\bigg\},
    \end{align*}
{where we again have used the identity $\nabla_{\mathbf{s}}\textrm{e}^{i\mathbf{p}\cdot\mathbf{s}/\hbar}=\frac{i\mathbf{p}}{\hbar}\textrm{e}^{i\mathbf{p}\cdot\mathbf{s}/\hbar}$. To evaluate the various inverse Laplace transforms above, we will return to the identity of \cref{invLap1} using $\xi=0,-1$, and $-2$ this time, and find
\begin{align*}
    A_{2}^{WK}=& -\frac{g\hbar^{2}}{12m}\bigg(\frac{m}{2\pi\hbar^{2}}\bigg)\big(\nabla_{\mathbf{r}}^{2}V(\mathbf{r})\big)\bigg(\frac{\hbar^{2}k_{F}}{m s}\bigg)^{d/2-2}J_{\frac{d}{2}-2}(z) \\
    &+\frac{g\hbar^{2}}{24m}\bigg(\frac{m}{2\pi\hbar^{2}}\bigg)\big(\nabla_{\mathbf{r}}V(\mathbf{r})\big)^{2}\bigg(\frac{\hbar^{2}k_{F}}{m s}\bigg)^{d/2-3}J_{\frac{d}{2}-3}(z)\\
    &+\frac{g}{8}\bigg(\frac{m}{2\pi\hbar^{2}}\bigg)\bigg(\nabla_{\mathbf{r}}V(\mathbf{r})\cdot\frac{\mathbf{s}}{s}\bigg)^{2}\bigg(\frac{\hbar^{2}k_{F}}{m s}\bigg)^{d/2-2}J_{\frac{d}{2}-2}(z) \\
    &-\frac{g}{6}\bigg(\frac{m}{2\pi\hbar^{2}}\bigg)\bigg[\nabla_{\mathbf{r}}\bigg(\nabla_{\mathbf{r}}V(\mathbf{r})\cdot\mathbf{s}\bigg)\cdot\mathbf{s}\bigg]\bigg(\frac{\hbar^{2}k_{F}}{m s}\bigg)^{d/2-1}J_{\frac{d}{2}-1}(z).
    \end{align*}
Similarly to before, eliminating the potential $V(\mathbf{r})$ in favor of $k_{F}^{2}$ through the use of $\nabla_{\mathbf{r}}V(\mathbf{r})=-\frac{\hbar^{2}}{2m}\nabla_{\mathbf{r}}k_{F}^{2}$ allows us to arrange the above in the final form}  
    \begin{align*}
    A_{2}^{WK}&= \frac{g k_{F}^{d-4}z^{2}}{24(2\pi z)^{d/2}}J_{\frac{d}{2}-2}(z)\big(\nabla_{\mathbf{r}}^{2}k_{F}^{2}\big)+\frac{g k_{F}^{d-6}z^{3}}{96(2\pi z)^{d/2}}J_{\frac{d}{2}-3}(z)\big(\nabla_{\mathbf{r}}k_{F}^{2}\big)^{2}\\
    &\quad+\frac{g k_{F}^{d-6}z^{4}}{32(2\pi z)^{d/2}}J_{\frac{d}{2}-2}(z)\bigg(\nabla_{\mathbf{r}}k_{F}^{2}\cdot\frac{\mathbf{s}}{s}\bigg)^{2}+\frac{g k_{F}^{d-4}z^{3}}{12(2\pi z)^{d/2}}J_{\frac{d}{2}-1}(z)\bigg[\nabla_{\mathbf{r}}\bigg(\nabla_{\mathbf{r}}k_{F}^{2}\cdot\frac{\mathbf{s}}{s}\bigg)\cdot\frac{\mathbf{s}}{s}\bigg],
    \end{align*}
{an expression identical to \cref{A2term}, which allows us to write
\begin{gather*}
    A_{2}^{WK}=A_{2}^{K}.
\end{gather*}
Finally, combining the above results for $A_{0,1,2}^{WK}$, we have proved the equivalence
\begin{gather*}
    \rho_{WK}(\mathbf{r},\mathbf{r}')=\rho_{K}(\mathbf{r},\mathbf{r}').
\end{gather*}}

\section{Useful relationships}\label{appendix:relations}
\subsection{Bessel function identities}

In this subsection, we list specific identities of cylindrical Bessel functions $J_{\nu}(z)$ of order $\nu$ which are used throughout our work~\cite{gradshteintable2015},
    \begin{align}
        J_{\nu}(z)&=\frac{2(\nu+1)}{z}J_{\nu+1}(z)-J_{\nu+2}(z),    \label{Bessel1} \\
        J_{\nu}(z)&=\frac{\big(4(\nu+1)(\nu+2)-z^{2}\big)J_{\nu+2}(z)-2z(\nu+1)J_{\nu+3}(z)}{z^{2}},    \label{Bessel2} \\
        \frac{\partial(z^{-\nu}J_{\nu}(z))}{\partial z}&=-z^{-\nu}J_{\nu+1}(z).\label{Bessel3} 
    \end{align}

\subsection{Inverse Laplace transform identities}

In this subsection, we evaluate the following inverse Laplace transformation which will be used in various places throughout our work.
    \begin{align}
        \mathcal{L}_{\mu}^{-1}\bigg\{\beta^{-d/2-\xi}\textrm{e}^{-\frac{m\mathbf{s}^{2}}{2\hbar^{2}\beta}-\beta V(\mathbf{R})}\bigg\}
        &=\int_{0}^{\mu}\bigg(\frac{2\hbar^{2}(\mu-\mu')}{m \mathbf{s}^{2}}\bigg)^{(d/2+\xi-1)/2} \nonumber\\
        &\quad\times J_{\frac{d}{2}+\xi-1}\Bigg(\sqrt{\frac{2m\mathbf{s}^{2}}{\hbar^{2}}(\mu-\mu')}\Bigg)\delta\big(\mu'-V(\mathbf{R})\big)d\mu' \nonumber \\
        &=\bigg(\frac{2\hbar^{2}\big(\mu-V(\mathbf{R})\big)}{m \mathbf{s}^{2}}\bigg)^{(d/2+\xi-1)/2}\nonumber\\
        &\quad\times J_{\frac{d}{2}+\xi-1}\Bigg(\sqrt{\frac{2m\mathbf{s}^{2}}{\hbar^{2}}\big(\mu-V(\mathbf{R})\big)}\Bigg)\nonumber \\
        &=\Bigg(\frac{\hbar^{4}k_{F}^{2}}{m^{2} \mathbf{s}^{2}}\Bigg)^{(d/2+\xi-1)/2} J_{\frac{d}{2}+\xi-1}(z) \label{invLap1}, 
    \end{align}
where $\xi$ is some real constant which, in this paper, only ever takes values $\xi=1,0,-1,-2$, while $\beta$ {(}not to be confused with the inverse of the temperature{)} and the Fermi energy $\mu$ are conjugate variables of the Laplace transformation. 
Going between the first and second lines in the derivation of \cref{invLap1}, we utilized three common inverse Laplace transforms which are listed below~\cite{bracksemiclassical2018,gradshteintable2015}, and going between the third and fourth lines we made the substitution $\mu-V(\mathbf{R})=\frac{\hbar^{2}}{2 m}k_{F}(\mathbf{R})^{2}$, suppressed the $\mathbf{R}$ dependent notation of $k_{F}$ so that $k_{F}=k_{F}(\mathbf{R})$, and defined $z=k_{F} s$.
    \begin{align*}
        \mathcal{L}_{\mu}^{-1}\big\{\textrm{e}^{-\beta V(\mathbf{R})}\big\}&=\delta\big(\mu-V(\mathbf{R})\big), \\
        \mathcal{L}_{\mu}^{-1}\big\{F(\beta)G(\beta)\big\}&=\int_{0}^{\mu}f(\mu-\mu')g(\mu')d\mu', \\
        \mathcal{L}_{\mu}^{-1}\bigg\{\frac{1}{\beta^{d/2+\xi}}\textrm{e}^{-\frac{m\mathbf{s}^{2}}{2\hbar^{2}\beta}}\bigg\}&=\bigg(\frac{2\hbar^{2}\mu}{m \mathbf{s}^{2}}\bigg)^{(d/2+\xi-1)/2}J_{\frac{d}{2}+\xi-1}\Bigg(\sqrt{\frac{2m\mathbf{s}^{2}}{\hbar^{2}}\mu}\Bigg) \nonumber.
    \end{align*}
$F(\beta)$ and $G(\beta)$ are arbitrary functions representing the Laplace transforms of the functions $f(\mu)$ and $g(\mu)$, respectively.
The third of the above three common inverse Laplace transform only holds true when $\frac{m\mathbf{s}^{2}}{2\hbar^{2}}>-1$.
In our case, this is always true since this combination of values is in fact always greater than $0$.
We also emphasize that \cref{invLap1} is utilized in \ref{appendix:WKDOM} with $\mathbf{R}$ replaced by $\mathbf{r}$.

\subsection{Gradient identities}

In this subsection we list some exact and approximate identities between both gradients and derivatives with respect to $\mathbf{r}$, $\mathbf{R}$, $\mathbf{s}$, and $z$. These are applied in various places throughout our work and are given by the following expressions:
    \begin{align}
        \nabla_{\mathbf{r}}&=\frac{1}{2}\nabla_{\mathbf{R}}+\nabla_{\mathbf{s}} \label{gradrA},\\
         \nabla_{\mathbf{r}}^{2}&=\frac{1}{4}\nabla_{\mathbf{R}}^{2}+\nabla_{\mathbf{s}}^{2}+\nabla_{\mathbf{R}}\cdot\nabla_{\mathbf{s}},  \label{gradrB}\\
        \nabla_{\mathbf{s}}&=k_{F}\frac{\mathbf{s}}{s}\frac{\partial}{\partial z}, \label{gradS}\\
        \nabla_{\mathbf{r}}V(\mathbf{r})&\approx\nabla_{\mathbf{R}}V(\mathbf{R})+\frac{1}{2}\nabla_{\mathbf{R}}\big(\nabla_{\mathbf{R}}V(\mathbf{R})\cdot\mathbf{s}\big) \label{gradr1}, \\
        \nabla_{\mathbf{r}}^{2}V(\mathbf{r})&\approx\nabla_{\mathbf{R}}^{2}V(\mathbf{R}) \label{gradr2}, \\
        \nabla_{\mathbf{r}}\bigg(\nabla_{\mathbf{r}}V(\mathbf{r})\cdot\frac{\mathbf{s}}{s}\bigg)\cdot\frac{\mathbf{s}}{s}&{\approx} \nabla_{\mathbf{R}}\bigg(\nabla_{\mathbf{R}}V(\mathbf{R})\cdot\frac{\mathbf{s}}{s}\bigg)\cdot\frac{\mathbf{s}}{s}\label{gradr3}. 
    \end{align}
As is the case throughout {this} paper, an equal ($=$) sign denotes an exact identity, and an approximation sign ($\approx$) denotes an approximate one where only terms up to and including second order in { derivatives of the potential $V$ (equivalently $\hbar$) are retained.} \Cref{gradrA,gradrB,gradS} are straightforward to show via \cref{symmCoord} and the definition of $z$, but \cref{gradr1,gradr2,gradr3} are more complicated and deserve their own proofs.
{To prove \cref{gradr1} we use \cref{gradrA} and the Taylor expansion up to second order in relative coordinates $\mathbf{s}$ of $V(\mathbf{r})=V(\mathbf{R}+\mathbf{s}/2)$, given by \cref{Vseries}. We then have}
\begin{align}
        \nabla_{\mathbf{r}}V(\mathbf{r})=\nabla_{\mathbf{r}}V(\mathbf{R}+\mathbf{s}/2){=}&{\frac{1}{2}\nabla_{\mathbf{R}}V(\mathbf{R}+\mathbf{s}/2)+\nabla_{\mathbf{s}}V(\mathbf{R}+\mathbf{s}/2)} \nonumber\\
        =&\frac{1}{2}\nabla_{\mathbf{R}}\bigg(V(\mathbf{r})+\frac{\nabla_{\mathbf{R}}V(\mathbf{R})\cdot\mathbf{s}}{2}+\frac{1}{8}\Big[\nabla_{\mathbf{R}}\big(\nabla_{\mathbf{R}}V(\mathbf{R})\cdot\mathbf{s}\big)\cdot\mathbf{s}\Big]\bigg) \nonumber \\
        &+\nabla_{\mathbf{s}}\bigg(V(\mathbf{r})+\frac{\nabla_{\mathbf{R}}V(\mathbf{R})\cdot\mathbf{s}}{2}+\frac{1}{8}\Big[\nabla_{\mathbf{R}}\big(\nabla_{\mathbf{R}}V(\mathbf{R})\cdot\mathbf{s}\big)\cdot\mathbf{s}\Big]\bigg). \nonumber
    \end{align}
{Taking the above expression and only keeping terms up to second derivatives of the potential $V$, we obtain
\begin{align}
        \nabla_{\mathbf{r}}V(\mathbf{r})\approx&\frac{1}{2}\nabla_{\mathbf{R}}\bigg(V(\mathbf{r})+\frac{\nabla_{\mathbf{R}}V(\mathbf{R})\cdot\mathbf{s}}{2}\bigg)+\nabla_{\mathbf{s}}\bigg(\frac{\nabla_{\mathbf{R}}V(\mathbf{R})\cdot\mathbf{s}}{2}+\frac{1}{8}\Big[\nabla_{\mathbf{R}}\big(\nabla_{\mathbf{R}}V(\mathbf{R})\cdot\mathbf{s}\big)\cdot\mathbf{s}\Big]\bigg) \nonumber \\
        =&\frac{1}{2}\nabla_{\mathbf{R}}V(\mathbf{R})+\frac{1}{4}\nabla_{\mathbf{R}}\big(\nabla_{\mathbf{R}}V(\mathbf{R})\cdot\mathbf{s}\big)\nonumber\\
        &+\frac{1}{2}\nabla_{\mathbf{s}}\big(\nabla_{\mathbf{R}}V(\mathbf{R})\cdot\mathbf{s}\big)+\frac{1}{8}\nabla_{\mathbf{s}}\Big[\nabla_{\mathbf{R}}\big(\nabla_{\mathbf{R}}V(\mathbf{R})\cdot\mathbf{s}\big)\cdot\mathbf{s}\Big] \nonumber \\
        =& \frac{1}{2}\nabla_{\mathbf{R}}V(\mathbf{R})+\frac{1}{4}\nabla_{\mathbf{R}}\big(\nabla_{\mathbf{R}}V(\mathbf{R})\cdot\mathbf{s}\big)+\frac{1}{2}\nabla_{\mathbf{R}}V(\mathbf{R})+\frac{1}{4}\nabla_{\mathbf{R}}\big(\nabla_{\mathbf{R}}V(\mathbf{R})\cdot\mathbf{s}\big) \nonumber \\
        {\color{black}=}& {\color{black}\nabla_{\mathbf{R}}V(\mathbf{R})+\frac{1}{2}\nabla_{\mathbf{R}}\big(\nabla_{\mathbf{R}}V(\mathbf{R})\cdot\mathbf{s}\big)} \nonumber.     
    \end{align}}
{Next, to derive \cref{gradr2} we start from}
    \begin{align}
        \nabla_{\mathbf{r}}^{2}V(\mathbf{r})&=\nabla_{\mathbf{r}}^{2}V(\mathbf{R}+\mathbf{s}/2) \nonumber \\
        {=}&{\frac{1}{4}\nabla_{\mathbf{R}}^{2}V(\mathbf{R}+\mathbf{s}/2)+\nabla_{\mathbf{s}}^{2}V(\mathbf{R}+\mathbf{s}/2)+(\nabla_{\mathbf{R}}\cdot\nabla_{\mathbf{s}})V(\mathbf{R}+\mathbf{s}/2)} \nonumber,
    \end{align}
{where we have used \cref{gradrB}. Applying the expansion of \cref{Vseries} in a similar way as was done in the derivation of \cref{gradr1}, we neglect terms with higher derivatives than second order of the potential $V$ and reduce the above expression to}
    \begin{align}
        \nabla_{\mathbf{r}}^{2}V(\mathbf{r})&\approx \frac{1}{4}\nabla_{\mathbf{R}}^{2}V(\mathbf{R})+\frac{1}{8}\nabla_{\mathbf{s}}^{2}\Big[\nabla_{\mathbf{R}}\big(\nabla_{\mathbf{R}}V(\mathbf{R})\cdot\mathbf{s}\big)\cdot\mathbf{s}\Big]+\frac{1}{2}\big(\nabla_{\mathbf{R}}\cdot\nabla_{\mathbf{s}}\big)\big(\nabla_{\mathbf{R}}V(\mathbf{R})\cdot\mathbf{s}\big) \nonumber\\
        &= {\frac{1}{4}\nabla_{\mathbf{R}}^{2}V(\mathbf{R})+\frac{1}{4}\nabla_{\mathbf{R}}^{2}V(\mathbf{R})+\frac{1}{2}\nabla_{\mathbf{R}}^{2}V(\mathbf{R})} \nonumber\\
        &= \nabla_{\mathbf{R}}^{2}V(\mathbf{R}). \nonumber
    \end{align}
{To derive} \cref{gradr3} {we make use of \cref{gradrA,gradr1}, and again only retain terms up to second order in $\nabla_{\mathbf{R}}$ to find}
    \begin{align}
        \nabla_{\mathbf{r}}\bigg(\nabla_{\mathbf{r}}V(\mathbf{r})\cdot\frac{\mathbf{s}}{s}\bigg)\cdot\frac{\mathbf{s}}{s}
        {\approx}&{\bigg(\frac{1}{2}\nabla_{\mathbf{R}}+\nabla_{\mathbf{s}}\bigg)\bigg(\nabla_{\mathbf{R}}V(\mathbf{R})\cdot\frac{\mathbf{s}}{s}+\frac{1}{2}\nabla_{\mathbf{R}}\big(\nabla_{\mathbf{R}}V(\mathbf{R})\cdot\mathbf{s}\big)\cdot\frac{\mathbf{s}}{s}\bigg)\cdot\frac{\mathbf{s}}{s}} \nonumber \\ 
        {\approx}&{\frac{1}{2}\nabla_{\mathbf{R}}\bigg(\nabla_{\mathbf{R}}V(\mathbf{R})\cdot\frac{\mathbf{s}}{s}\bigg)\cdot\frac{\mathbf{s}}{s} +\nabla_{\mathbf{s}}\bigg(\nabla_{\mathbf{R}}V(\mathbf{R})\cdot\frac{\mathbf{s}}{s}\bigg)\cdot\frac{\mathbf{s}}{s}}\nonumber\\
        &{+\frac{1}{2}\nabla_{\mathbf{s}}\Big[\nabla_{\mathbf{R}}\big(\nabla_{\mathbf{R}}V(\mathbf{R})\cdot\mathbf{s}\big)\cdot\frac{\mathbf{s}}{s}\Big]\cdot\frac{\mathbf{s}}{s} \label{proof3a},}
    \end{align}
{ To further simplify \cref{proof3a} into \cref{gradr3}, we show that the second term vanishes identically by writing this term as}
{\begin{align}
            \nabla_{\mathbf{s}}\bigg(\nabla_{\mathbf{R}}V(\mathbf{R})\cdot\frac{\mathbf{s}}{s}\bigg)\cdot\frac{\mathbf{s}}{s}&=\sum_{i=1}^{d}\sum_{j=1}^{d}\frac{s_{i}}{s}\frac{\partial}{\partial s_{i}}\bigg(\frac{s_{j}}{s}\frac{\partial V(\mathbf{R})}{\partial R_{j}}\bigg)=\sum_{i=1}^{d}\sum_{j=1}^{d}\delta_{ij}\frac{s_{i}}{s^{2}}\frac{\partial V(\mathbf{R})}{\partial R_{j}}-\frac{s_{i}^{2}s_{j}}{s^{4}}\frac{\partial V(\mathbf{R})}{\partial R_{j}}\nonumber \\
            &=\sum_{i=1}^{d}\frac{s_{i}}{s^{2}}\frac{\partial V(\mathbf{R})}{\partial R_{i}}-\sum_{j=1}^{d}\frac{s_{j}}{s^{2}}\frac{\partial V(\mathbf{R})}{\partial R_{j}}=0 \label{proof3b},
\end{align}
while the third term of \cref{proof3a} simplifies to
\begin{align}
            \frac{1}{2}\nabla_{\mathbf{s}}\Big[\nabla_{\mathbf{R}}\big(\nabla_{\mathbf{R}}V(\mathbf{R})\cdot\mathbf{s}\big)\cdot\frac{\mathbf{s}}{s}\Big]\cdot\frac{\mathbf{s}}{s}&=\sum_{i=1}^{d}\sum_{j=1}^{d}\sum_{l=1}^{d}\frac{s_{i}}{2s}\frac{\partial}{\partial s_{i}}\bigg(\frac{s_{j}s_{l}}{s}\frac{\partial^{2}V(\mathbf{R})}{\partial R_{j}\partial R_{l}}\bigg)\nonumber \\
            &=\sum_{i=1}^{d}\sum_{j=1}^{d}\sum_{l=1}^{d}\delta_{ij}\frac{s_{i}s_{l}}{2s^{2}}\frac{\partial^{2}V(\mathbf{R})}{\partial R_{j}\partial R_{l}} \nonumber\\
            &\quad+\delta_{il}\frac{s_{i}s_{j}}{2s^{2}}\frac{\partial^{2}V(\mathbf{R})}{\partial R_{j}\partial R_{l}}-\frac{s_{i}^{2}s_{j}s_{l}}{2s^{4}}\frac{\partial^{2}V(\mathbf{R})}{\partial R_{j}\partial R_{l}} \nonumber\\
            &=\sum_{i=1}^{d}\sum_{l=1}^{d}\frac{s_{i}s_{l}}{2s^{2}}\frac{\partial^{2}V(\mathbf{R})}{\partial R_{i}\partial R_{l}}\nonumber\\
            &\quad+\sum_{i=1}^{d}\sum_{j=1}^{d}\frac{s_{i}s_{j}}{2s^{2}}\frac{\partial^{2}V(\mathbf{R})}{\partial R_{j}\partial R_{i}}-\sum_{j=1}^{d}\sum_{l=1}^{d}\frac{s_{j}s_{l}}{2s^{2}}\frac{\partial^{2}V(\mathbf{R})}{\partial R_{j}\partial R_{l}}\nonumber \\
            &=\sum_{i=1}^{d}\sum_{l=1}^{d}\frac{s_{i}s_{l}}{2s^{2}}\frac{\partial^{2}V(\mathbf{R})}{\partial R_{i}\partial R_{l}} =\frac{1}{2}\nabla_{\mathbf{R}}\bigg(\nabla_{\mathbf{R}}V(\mathbf{R})\cdot\frac{\mathbf{s}}{s}\bigg)\cdot\frac{\mathbf{s}}{s} \label{proof3c}.
\end{align}
Combining \cref{proof3a,proof3b,proof3c} finally yields \cref{gradr3}.}

{\section{Alternate derivation of WKODM, KODM, and GVODM equivalence}\label{appendix:alternate}}

\subsection{Symmetric Bloch matrix}
In references~\cite{redjatisemiclassical2019,bencheikhhermitian2016,bencheikhmanifestly2016}, the manifestly Hermitian and idempotent GVODM in symmetric coordinates is derived for $d=1,2,3,4$-dimensions despite the fact that it is through means of a non-Hermitian and non-idempotent WKODM method in terms of non-symmetric coordinates.
{In Section \ref{sec:KODM} we also proved the equivalence between the WKODM, KODM, and GVODM by applying the change of coordinates $(\mathbf{r},\mathbf{r}')\rightarrow (\mathbf{R},\mathbf{s})$ to the non-symmetric Wigner-Kirkwood expansion of the Bloch density matrix. For completeness, here} we will provide a{n additional yet equivalent} {$d$}-dimensional derivation {of the symmetric ODM equivalence which instead starts by symmetrizizing the definition of the Bloch density matrix, and then by performing the Wigner-Kirkwood expansion, now in symmetric coordinates.} Starting from the same point that reference~\cite{bracksemiclassical2018} {does} for their non-symmetric WKODM derivation, see their section 4.3{,} the non-symmetric Bloch {density }matrix is given by
    \begin{gather}
        C(\mathbf{r},\mathbf{r}';\beta)=g\langle\mathbf{r}\vert \textrm{e}^{-\beta\hat{H}}\vert \mathbf{r}'\rangle \label{blochMat}.
    \end{gather}
Note that one can easily derive the presence of the degeneracy factor $g$ by accounting for fermionic spin within the position states and simplifying. $\hat{H}$ is the quantum one-body Hamiltonian, and we also note that the WKODM {expansion} has yet to be applied.
Before changing to the symmetric coordinates, consider the unitary translation operator $\hat{U}(\mathbf{a})={\textrm{e}^{-(i/\hbar)\hat{\mathbf{P}}\cdot\mathbf{a}}}$, where $\hat{\mathbf{P}}=-i\hbar\nabla_{\mathbf{R}}$ is the $\mathbf{R}$-space representation of the momentum operator $\hat{\mathbf{P}}\vert\mathbf{p}\rangle=\mathbf{p}\vert \mathbf{p}\rangle$, and $\mathbf{a}$ is an arbitrary position space vector.
$\hat{U}(\mathbf{a})$ acts on position space states and scalar functions $f$ as follows~\cite{sakuraimodern2021}
    \begin{align}
        \hat{U}(\mathbf{a})\vert \mathbf{R}\rangle&=\vert\mathbf{R}-\mathbf{a}\rangle, \nonumber \\
        \hat{U}(\mathbf{a})^{\dagger}\vert \mathbf{R}\rangle&=\vert\mathbf{R}+\mathbf{a}\rangle, \nonumber \\
        \hat{U}(\mathbf{a})^{\dagger}f(\mathbf{b})\hat{U}(\mathbf{a})&=f(\mathbf{a}+\mathbf{b}), \nonumber
    \end{align}
where $\mathbf{b}$ is an arbitrary position space vector.
Applying \cref{symmCoord} to \cref{blochMat} and using the above unitary identities gives the expression for the Bloch matrix in terms of symmetric coordinates by
    \begin{align}
        C(\mathbf{R},\mathbf{s};\beta)&=g\langle\mathbf{R}+\mathbf{s}/2\vert \textrm{e}^{-\beta\hat{H}}\vert \mathbf{R}-\mathbf{s}/2\rangle \nonumber \\
        &= g\langle\mathbf{R}\vert \textrm{e}^{-\frac{i}{\hbar}\hat{\mathbf{P}}\cdot\mathbf{s}/2}\textrm{e}^{-\beta\hat{H}}\textrm{e}^{-\frac{i}{\hbar}\hat{\mathbf{P}}\cdot\mathbf{s}/2}\vert \mathbf{R}\rangle \nonumber \\
        &= g\langle\mathbf{R}\vert \textrm{e}^{-\frac{i}{\hbar}\hat{\mathbf{P}}\cdot\mathbf{s}}\hat{U}(\mathbf{s}/2)^{\dagger}\textrm{e}^{-\beta\hat{H}}\hat{U}(\mathbf{s}/2)\vert \mathbf{R}\rangle \nonumber \\
        &= g\langle\mathbf{R}\vert \textrm{e}^{-\frac{i}{\hbar}\hat{\mathbf{P}}\cdot\mathbf{s}}\textrm{e}^{-\beta\hat{U}(\mathbf{s}/2)^{\dagger}\hat{H}\hat{U}(\mathbf{s}/2)}\vert \mathbf{R}\rangle \nonumber \\
        &= g\langle\mathbf{R}\vert \textrm{e}^{-\frac{i}{\hbar}\hat{\mathbf{P}}\cdot\mathbf{s}}\textrm{e}^{-\beta\frac{\hat{\mathbf{P}}^{2}}{2m}-\beta V(\mathbf{R}+\mathbf{s}/2)}\vert \mathbf{R}\rangle,\label{blochMat2}
    \end{align}
where in between the fourth and last lines we made the substitution $\hat{H}=\hat{\mathbf{P}}^{2}/(2m)+V(\mathbf{R})$.
Making use of the momentum-space completeness relationship $\mathds{1}=\int d^{d}p\;\vert \mathbf{p}\rangle\langle\mathbf{p}\vert$ and the inner product relationship $\langle\mathbf{R}\vert \mathbf{p}\rangle=(2\pi\hbar)^{-d/2}{\textrm{e}^{(i/\hbar)\mathbf{p}\cdot\mathbf{R}}}$ allows us to recast \cref{blochMat2} as
    \begin{gather}
        C(\mathbf{R},\mathbf{s};\beta)=\frac{g}{(2\pi\hbar)^{d/2}}\int d^{d}p\;\textrm{e}^{-\frac{i}{\hbar}\mathbf{p}\cdot\mathbf{s}}\textrm{e}^{\frac{i}{\hbar}\mathbf{p}\cdot\mathbf{R}}\langle\mathbf{p}\vert\textrm{e}^{-\beta\frac{\hat{\mathbf{P}}^{2}}{2m}-\beta V(\mathbf{R}+\mathbf{s}/2)}\vert \mathbf{R}\rangle \label{blochMat3}.
    \end{gather}
At this point, we can perform our symmetric semiclassical WKODM {method} following the same techniques used for the usual non-symmetric WKODM method within reference~\cite{bracksemiclassical2018}.
The process starts by making the substitution $\langle\mathbf{p}\vert{\textrm{e}^{-\beta\frac{\hat{\mathbf{P}}^{2}}{2m}-\beta V(\mathbf{R}+\mathbf{s}/2)}}\vert \mathbf{R}\rangle={\textrm{e}^{-\beta H_{cl}(\mathbf{R}+\mathbf{s}/2,\mathbf{p})}}\langle\mathbf{p}\vert\mathbf{R}\rangle\tilde{w}(\mathbf{R},\mathbf{s},\mathbf{p};\beta)$, {where we recall from before that $H_{cl}(\mathbf{r},\mathbf{p})=\frac{\mathbf{p}^{2}}{2m}+V(\mathbf{r})$. This yields}
    \begin{gather}
        C(\mathbf{R},\mathbf{s};\beta)=\frac{g}{(2\pi\hbar)^{d}}\int d^{d}p\;\textrm{e}^{-\frac{i}{\hbar}\mathbf{p}\cdot\mathbf{s}}\textrm{e}^{-\beta H_{cl}(\mathbf{R}+\mathbf{s}/2,\mathbf{p})}\tilde{w}(\mathbf{R},\mathbf{s},\mathbf{p};\beta), \label{blochMat4}
    \end{gather}
where $\tilde{w}$ is a function which obeys 
    \begin{gather}
        \textrm{e}^{-\beta H_{cl}(\mathbf{R}+\mathbf{s}/2,{\mathbf{p}})-(i/\hbar)\mathbf{p}\cdot\mathbf{R}}\tilde{w}(\mathbf{R},\mathbf{s},\mathbf{p};\beta)=\tilde{u}(\mathbf{R},\mathbf{s},\mathbf{p};\beta), \label{uDef}
    \end{gather}
and $\tilde{u}(\mathbf{R},\mathbf{s},\mathbf{p};\beta)$ is a function that obeys the Bloch equation {( as defined by equation (4.35) in reference ~\cite{bracksemiclassical2018})}
    \begin{gather}
        \frac{\partial \tilde{u}}{\partial\beta}+\hat{H}\tilde{u}=0. \label{blochEqn}
    \end{gather}
{Here the Hamiltonian takes its position space representation, $\hat{H}=-\hbar^{2}\nabla_{\mathbf{R}}^{2}/(2m)+V(\mathbf{R})$, while $\tilde{u}(\mathbf{R},\mathbf{s},\mathbf{p};\beta)$ treats both the momentum $\mathbf{p}$ and position coordinates $\mathbf{R}$ and $\mathbf{s}$ on the same footing as classical variables (opposed to operators).} For simplicity, we sometimes suppress the $(\mathbf{R},\mathbf{s},\mathbf{p};\beta)$ dependence of both the $\tilde{w}$ and $\tilde{u}$ functions, in a similar form to \cref{blochEqn}. 

\subsection{Symmetric WKODM expansion} 
Substituting \cref{uDef} into \cref{blochEqn}, making the second order semiclassical approximation $\tilde{w}(\mathbf{R},\mathbf{s},\mathbf{p};\beta)\approx1+\hbar \tilde{w}_{1}(\mathbf{R},\mathbf{s},\mathbf{p};\beta)+\hbar^{2}\tilde{w}_{2}(\mathbf{R},\mathbf{s},\mathbf{p};\beta)$, and collecting terms based on their order of $\hbar$ {while dividing through by a common factor of $\textrm{e}^{-\beta H_\text{cl}(R+s/2,p)-(i/\hbar) p\cdot R}$} yields
    \begin{gather}
        \hbar\bigg(-\frac{i\beta}{m}\mathbf{p}\cdot\nabla_{\mathbf{R}}V(\mathbf{R}+\mathbf{s}/2)+\frac{\partial \tilde{w}_{1}}{\partial \beta}\bigg)+\hbar^{2}\bigg(-\frac{i\beta}{m}\mathbf{p}\cdot\nabla_{\mathbf{R}}V(\mathbf{R}+\mathbf{s}/2)\tilde{w}_{1} \nonumber \\
        -\frac{\beta^{2}}{2m}\big(\nabla_{\mathbf{R}}V(\mathbf{R}+\mathbf{s}/2)\big)^{2}+\frac{\beta}{2m}\nabla_{\mathbf{R}}^{2}V(\mathbf{R}+\mathbf{s}/2)
        +\frac{i}{{m}}(\mathbf{p}\cdot\nabla_{\mathbf{R}})\tilde{w}_{1}+\frac{{\partial}\tilde{w}_{2}}{{\partial\beta}}\bigg)=0. \label{blochEqn2}
    \end{gather}
In order for this equation to be true, the expressions within the large brackets multiplying both $\hbar$ and $\hbar^{2}$ must both be equal to zero.
The expression multiplying $\hbar$ can be easily be solved to obtain $\tilde{w}_{1}$, which can then be substituted into the expression multiplying $\hbar^{2}$ to solve for $\tilde{w}_{2}$.
Doing this yields
    \begin{align}
        \tilde{w}_{1}(\mathbf{R},\mathbf{s},\mathbf{p};\beta)
        &=\bigg(\frac{i\beta^{2}}{2m}\bigg)\big({\nabla_{\mathbf{R}}V(\mathbf{R}+\mathbf{s}/2)\cdot\mathbf{p}}\big), \label{w1Tilde}\\
        \tilde{w}_{2}(\mathbf{R},\mathbf{s},\mathbf{p};\beta)
        &=-\frac{\beta^{{2}}}{4m}\nabla_{\mathbf{R}}^{2}V(\mathbf{R}+\mathbf{s}/2) +\frac{\beta^{3}}{6m}\big(\nabla_{\mathbf{R}}V(\mathbf{R}+\mathbf{s}/2)\big)^{2} \nonumber \\
        &\quad{-\frac{\beta^{4}}{8m^{2}}\big(\nabla_{\mathbf{R}}V(\mathbf{R}+\mathbf{s}/2)\cdot\mathbf{p}\big)^{2}+\frac{\beta^{3}}{6m^{2}}(\mathbf{p}\cdot\nabla_{\mathbf{R}})^{2}V(\mathbf{R}+\mathbf{s}/2)}. \label{w2Tilde}
    \end{align}
Aside from a minus sign difference in \cref{w1Tilde}, \cref{w1Tilde,w2Tilde} are the same in form as their non-symmetric counterparts $w_{1}(\mathbf{r},\mathbf{r}';\beta)$ and $w_{2}(\mathbf{r},\mathbf{r}';\beta)$ respectively, as shown in {reference}~\cite{bracksemiclassical2018}.
To keep our notation consistent with the non-symmetric WKODM method, we utilize the fact that \cref{blochMat4} remains invariant when the integrand undergoes the momentum transformation $\mathbf{p}\rightarrow -\mathbf{p}$ (this transformation does not include changing the sign of the differential element $d^{d}p$).
This is shown below by \cref{intIdent}, where $f(\mathbf{p})$ is any function which is either a polynomial or can {be} expressed as a polynomial by a Taylor expansion.
    \begin{gather}
        \int d^{d}p\;\textrm{e}^{-\frac{i}{\hbar}\mathbf{p}\cdot\mathbf{s}-\beta\frac{\mathbf{p}^{2}}{2m}}f(\mathbf{p})=\int d^{d}p\;\textrm{e}^{-\frac{i}{\hbar}(-\mathbf{p})\cdot\mathbf{s}-\beta\frac{(-\mathbf{p})^{2}}{2m}}f(-\mathbf{p}) \label{intIdent}
    \end{gather}
Applying \cref{intIdent} allows us to re-express \cref{blochMat4} as
    \begin{align}
        C(\mathbf{R},\mathbf{s};\beta)&=\frac{g}{(2\pi\hbar)^{d}}\int d^{d}p\;\textrm{e}^{\frac{i}{\hbar}\mathbf{p}\cdot\mathbf{s}}\textrm{e}^{-\beta H_{cl}(\mathbf{R}+\mathbf{s}/2,\mathbf{p})}\tilde{w}(\mathbf{R},\mathbf{s},-\mathbf{p};\beta) \nonumber \\
        &=\frac{g}{(2\pi\hbar)^{d}}\int d^{d}p\;\textrm{e}^{\frac{i}{\hbar}\mathbf{p}\cdot\mathbf{s}}\textrm{e}^{-\beta H_{cl}(\mathbf{R}+\mathbf{s}/2,\mathbf{p})}w(\mathbf{R},\mathbf{s},\mathbf{p};\beta), \label{blochMat5}
    \end{align}
where we have defined $w(\mathbf{R},\mathbf{s},\mathbf{p};\beta)=\tilde{w}(\mathbf{R},\mathbf{s},-\mathbf{p};\beta)${. Applying the semiclassical expansion from before, $\tilde{w}(\mathbf{R},\mathbf{s},\mathbf{p};\beta)\approx1+\hbar \tilde{w}_{1}(\mathbf{R},\mathbf{s},\mathbf{p};\beta)+\hbar^{2}\tilde{w}_{2}(\mathbf{R},\mathbf{s},\mathbf{p};\beta)$ and \cref{w1Tilde,w2Tilde}, we can approximate \cref{blochMat5} by making the substitution $w(\mathbf{R},\mathbf{s},\mathbf{p};\beta)\approx 1+\hbar w_{1}(\mathbf{R},\mathbf{s},\mathbf{p};\beta)+\hbar^{2}w_{2}(\mathbf{R},\mathbf{s},\mathbf{p};\beta)$, where}
   \begin{align}
        w_{1}(\mathbf{R},\mathbf{s},\mathbf{p};\beta)&=\tilde{w}_{1}(\mathbf{R},\mathbf{s},-\mathbf{p};\beta)=-\bigg(\frac{i\beta^{2}}{2m}\bigg)\big({\nabla_{\mathbf{R}}V(\mathbf{R}+\mathbf{s}/2)\cdot\mathbf{p}}\big) \label{w1},\\
        w_{2}(\mathbf{R},\mathbf{s},\mathbf{p};\beta)&=\tilde{w}_{2}(\mathbf{R},\mathbf{s},-\mathbf{p};\beta) \nonumber\\
        &=-\frac{\beta^{{2}}}{4m}\nabla_{\mathbf{R}}^{2}V(\mathbf{R}+\mathbf{s}/2) +\frac{\beta^{3}}{6m}\big(\nabla_{\mathbf{R}}V(\mathbf{R}+\mathbf{s}/2)\big)^{2} \nonumber \\
        &\quad{-\frac{\beta^{4}}{8m^{2}}\big(\nabla_{\mathbf{R}}V(\mathbf{R}+\mathbf{s}/2)\cdot\mathbf{p}\big)^{2}+\frac{\beta^{3}}{6m^{2}}(\mathbf{p}\cdot\nabla_{\mathbf{R}})^{2}V(\mathbf{R}+\mathbf{s}/2).} \label{w2}
    \end{align}
{} \Cref{w1,w2} are {respectively} in the same form as the non-symmetric expressions of $w_{1}(\mathbf{r},\mathbf{r}';\beta)$ and $w_{2}(\mathbf{r},\mathbf{r}';\beta)$, given {in reference \cite{bracksemiclassical2018}}.
Similar to the procedure we followed both there and within Section~\ref{sec:KODM} for the KODM derivation (based on reference~\cite{bracksemiclassical2018}), we are finally in a position to derive the symmetric WKODM $\rho_{WK}(\mathbf{R},\mathbf{s})$ by utilizing inverse Laplace transforms and separating terms based on their order of $\hbar$.
We find {to second order}
   \begin{align}
    \rho_{WK}(\mathbf{R},\mathbf{s})&{=} B_{0}^{{WK}}+B_{1}^{{WK}}+B_{2}^{{WK}}; \label{WKODM} \\
       B_{0}^{{WK}}&=\mathcal{L}^{-1}_{\mu}\bigg\{\frac{C_{0}^{{WK}}(\mathbf{R},\mathbf{s};\beta)}{\beta}\bigg\} \label{B0term2}, \\
        B_{1}^{{WK}}&=\mathcal{L}^{-1}_{\mu}\bigg\{\frac{C_{1}^{{WK}}(\mathbf{R},\mathbf{s};\beta)}{\beta}\bigg\} \label{B1term2}, \\
        B_{2}^{{WK}}&=\mathcal{L}^{-1}_{\mu}\bigg\{\frac{C_{2}^{{WK}}(\mathbf{R},\mathbf{s};\beta)}{\beta}\bigg\} \label{B2term2}, 
    \end{align}
where each term of the approximate symmetric Bloch matrix $C^{{WK}}(\mathbf{R},\mathbf{s};\beta)\approx C_{0}^{{WK}}(\mathbf{R},\mathbf{s};\beta)+C_{1}^{{WK}}(\mathbf{R},\mathbf{s};\beta)+C_{2}^{{WK}}(\mathbf{R},\mathbf{s};\beta)$ is given by
   \begin{align}
        C_{0}^{{WK}}(\mathbf{R},\mathbf{s};\beta)&=\frac{g}{(2\pi\hbar)^{d}}\int d^{d}p\;\textrm{e}^{-\beta H_{cl}(\mathbf{R}+\mathbf{s}/2,\mathbf{p})+\frac{i}{\hbar}\mathbf{p}\cdot\mathbf{s}} \label{BC0},\\
        C_{1}^{{WK}}(\mathbf{R},\mathbf{s};\beta)&={\frac{g\hbar}{(2\pi\hbar)^{d}}}\int d^{d}p\;\textrm{e}^{-\beta H_{cl}(\mathbf{R}+\mathbf{s}/2,\mathbf{p})+\frac{i}{\hbar}\mathbf{p}\cdot\mathbf{s}}w_{1}(\mathbf{R},\mathbf{s},\mathbf{p};\beta)  \label{BC1},\\
        C_{2}^{{WK}}(\mathbf{R},\mathbf{s};\beta)&=\frac{g\hbar^{2}}{(2\pi\hbar)^{d}}\int d^{d}p\;\textrm{e}^{-\beta H_{cl}(\mathbf{R}+\mathbf{s}/2,\mathbf{p})+\frac{i}{\hbar}\mathbf{p}\cdot\mathbf{s}}w_{2}(\mathbf{R},\mathbf{s},\mathbf{p};\beta)  \label{BC2}.
    \end{align}
From hereon, the calculations and approximations we perform to fully simplify and combine \cref{w1,w2,WKODM,B0term2,B1term2,B2term2,BC0,BC1,BC2} are identical to those used in Section~\ref{sec:KODM} for the KODM method, hence to avoid redundancy, we will not explicitly show each step. After simplification one exactly finds that in symmetric coordinates $B_{0}^{{WK}}=A_{0}^{{WK}}$ [\cref{A0term4}], $B_{1}^{{WK}}=A_{1}^{{WK}}$ [\cref{A1term4}], and $B_{2}^{{WK}}=A_{2}^{{WK}}$ [\cref{A2term3}].
Therefore, combining these results together via \cref{WKODM}, we find that the symmetric WKODM is exactly identical to the symmetric KODM [\cref{GVODM}], which we know is exactly the {$d$}-dimensional GVODM originally proposed in reference~\cite{redjatisemiclassical2019}.

\section*{References}
\bibliographystyle{unsrt}
\bibliography{ref4}

\end{document}